# Exact solution of quantum dynamics of a cantilever coupling to a single trapped ultracold ion.

T. Liu[1], M. Feng[1,2] and K.L. Wang[1,3]

1. The School of Science, Southwest University of Science and Technology, Mianyang 621002,China
2. State Key Laboratory of Magnetic Resonance and Atomic and Molecular Physics, Wuhan Institute of Physics and Mathematics, Chinese Academy of Sciences, Wuhan 430070, China
3. The Department of Modern Physics, University of Science and Technology of China, Hefei 230026,China

**Abstract**   The quantum behavior of a precooled cantilever can be probed highly efficiently by electrostatically coupling to a trapped ultracold ion, in which a fast cooling of the cantilever down to the ground vibrational state is possible. Within a simple model with an ultracold ion coupled to a cantilever with only few vibrational quanta, we solve the dynamics of the coupling system by a squeezed-state expansion technique, and can in principle obtain the exact solution of the time-evolution of the coupling system in the absence of the rotating-wave approximation. Comparing to the treatment under the rotating-wave approximation, we present a more accurate description of the quantum behavior of the cantilever.

## I Introduction

With recent rapid progress in microfabrication technology, the quantum electromechanical system (QEMS), with nano- to micro-meter scale mechanical resonator electrostatically coupled to comparably sized electrical devices, has been drawn much attention [1]. Typical QEMSs include displacement detectors strongly coupled to high-quality cantilevers or doubly-clamped beams [2,3], or to single electron transistors [4,5]. With current state-of-the -art techniques, the quantum behavior of the QEMS can also be efficiently evidenced by the coupled electrical devices under the proper temperature and vacuum conditions. For example, detection of quantum superposition of position states of mechanical resonators has been available by using an electrostatically coupled Cooper-pair box [6]. In addition, it is expected that single phonon detection [7] and quantum tunneling of mechanical degrees of freedom [8] could be done by means of QEMS. Moreover, with the idea of QEMS, it is becoming an exciting research frontier to couple atomic systems to nano-mechanical devices for demonstrating some new phenomena in condensed matter physics [1,9].

Besides its own interest, QEMS is also playing more and more important role in the context of quantum information processing (QIP). We have noticed an idea [10] for QIP with coupled trapped ion to a high-quality nano-mechanical resonator. Under certain conditions, the two component subsystems interact and yield energy exchange, entanglement and quantum gating. This is a promising way towards large-scale QIP. Moreover, readout of individual qubits is very challenging in solid-state QIP, which is resulted from the difficulty of the single spin detection. A recent breakthrough in this respect is the detection of a single unpaired electron-spin on the surface of silicon dioxide by magnetic resonance force microscopy [11], which is actually a QEMS. Another QEMS – single electron transistor based on a fullerene [12] also provides a promising way toward readout of qubits in fullerene-based QIP [13].

In this paper, we focus on the dynamics of a coupling system involving a very small doubly clamped cantilever (- a nonomechanical resonator as mentioned in [10]) and a single trapped ultracold ion, which has originally been studied in [14]. Our work can also be applied to [10] since there are many things overlapping between [10] and [14]. One of the purposes in [14] is to highly efficiently probe the vibration of a cantilever precooled to 4 Kelvin, through an electrostatically coupled trapped ion initially cooled to the vibrational ground state, as shown in Fig. 1. Since the deviations of the cantilever and the ion from their equilibrium positions in the oscillation are comparable to the equilibrium separation between them, a second-order perturbative expansion of the Coulomb interaction term could yield the coupling of the position operators of the ion and the cantilever. Experimentally this can be done simply by applying an external bias voltage at an electrode of the cantilever to couple the motion of the cantilever to the vibrational state of the ion. Since the detection of the quantum vibration of the trapped ion is a sophisticated work by optical methods – radiation with blue or red detuned laser [15] followed by electric shelving amplification [16], if the two component subsystems can be coupled and decoupled at will by adjusting the bias gate voltage, the scheme in [14] is available with current technique by following three steps: The first step is to switch on the coupling of the cantilever to the ion. At a suitable time point when the vibration of the cantilever is fully transferred to that of the ion, the coupling is switched off and the red or blue detuned laser will be on to couple the vibrational and the internal degrees of freedom of the ion. The last step is to read out the vibrational quanta by detecting the internal levels of the ion.

The other purpose in [14] is to fast cool the cantilever down to the ground vibrational state. The numerical calculation in [14] has shown the highly efficient transfer of the phonon from the cantilever to the ion. So the cantilever reaches its ground vibrational state only after a half coupling period.

The focus of the present paper is to exactly solve, within the simple model given in [14], the dynamics of the coupling system consisting of a cantilever and an ion both under decay due to the thermal environment, by a squeezed-state expansion technique. In Section II, we will present the analytical solutions by our technique for the systems under the rotating-wave approximation (RWA) and beyond the RWA, respectively. The numerical results demonstrated in Section III show the big differences of the RWA solution and the non-RWA one in the case of relatively strong coupling, which implies our investigation in the absence of the RWA to be essential to the detection and to the cooling of the cantilever. Some discussions about the validity and the experimental feasibility of our theoretical work as well as the conclusion appear in Section IV, and we put some detailed deduction steps in Appendix parts.

## II The solution by squeezed-state expansion technique

Starting from [14], we consider the first step by switching on the coupling between the cantilever and the trapped ion, which yields the following Hamiltonian in units of $\hbar = 1$,

$$H = (\omega - i\Gamma_a)a^+ a + (\nu - i\Gamma_b)b^+ b - \kappa(a + a^+)(b + b^+), \quad (1)$$

where ω and υ are vibrational frequencies of the cantilever and the trapped ion, respectively, and κ is the coupling constant between the two vibational degrees of freedom. To make a general treatment, we remain the counter-rotating terms in Eq. (1). So our study from Eq. (1) could be made in both weak and strong coupling cases. $\Gamma_a$ and $\Gamma_b$ are the decay coefficients regarding the vibrations of the cantilever and the trapped ion, respectively. We assume a trial solution of the wavefunction

$$|t\rangle = \rho(t) e^{\alpha_1(t)a^+ a^+ + \alpha_2(t)a^+ b^+ + \alpha_3(t)b^+ b^+} |0\rangle = \rho(t) | A(t) \rangle, \quad (2)$$

where $\rho(t)$, $\alpha_1(t)$, $\alpha_2(t)$, and $\alpha_3(t)$ are variables determined later. With Eq. (2), the Schrödinger equation of Eq. (1) yields,

$$\begin{aligned} i[\dot{\rho}(t) &+ \rho(t)\dot{\alpha}_1(t)a^+ a^+ + \rho(t)\dot{\alpha}_2(t)a^+ b^+ + \rho(t)\dot{\alpha}_3(t)b^+ b^+] | A(t) \rangle \\ &= \rho(t)\{2(\omega - i\Gamma_a)\alpha_1(t)a^+ a^+ + (\omega - i\Gamma_a)\alpha_2(t)a^+ b^+ \\ &+ (\nu - i\Gamma_b)\alpha_2(t)a^+ b^+ + 2(\nu - i\Gamma_b)\alpha_3(t)b^+ b^+ \\ &- \kappa[\alpha_2(t) + [2\alpha_1(t)\alpha_2(t) + \alpha_2(t)]a^+ a^+ + [2\alpha_2(t)\alpha_3(t) + \alpha_2(t)]b^+ b^+ \\ &+ [1 + \alpha_2^2(t) + 4\alpha_1(t)\alpha_3(t) + 2\alpha_1(t) + 2\alpha_3(t)]a^+ b^+]\} | A(t) \rangle. \end{aligned} \quad (3)$$

Comparing the terms $|A(t)\rangle, a^+ a^+ |A(t)\rangle, a^+ b^+ |A(t)\rangle, b^+ b^+ |A(t)\rangle$ in both sides of the equation, we obtain a set of equations,

$$i\dot{\rho}(t) = -\kappa \alpha_2(t) \rho(t), \quad (4)$$

$$i\dot{\alpha}_1(t) = 2(\omega - i\Gamma_a)\alpha_1(t) - \kappa\alpha_2(t) - 2\kappa\alpha_2(t)\alpha_1(t), \quad (5)$$

$$\begin{aligned} i\dot{\alpha}_2(t) &= (\omega + \nu - i\Gamma_a - i\Gamma_b)\alpha_2(t) - \kappa - \kappa\alpha_2^2(t) \\ &- 4\kappa\alpha_1(t)\alpha_3(t) - 2\kappa\alpha_1(t) - 2\kappa\alpha_3(t)], \end{aligned} \quad (6)$$

$$i\dot{\alpha}_3(t) = 2(\nu - i\Gamma_b)\alpha_3(t) - \kappa\alpha_2(t) - 2\kappa\alpha_2(t)\alpha_3(t). \quad (7)$$

From the initial condition $\alpha_2(0) = 0, \alpha_3(0) = 0$, and $\langle t=0 | t=0 \rangle = 1$, we have

$$\frac{\rho(0)^2}{\sqrt{(1 - 4|\alpha_1(0)|^2)}} = 1, \quad (8)$$

where the proof of Eq. (8) is put in Appendix I. Another initial condition is for the initial vibrational population of the cantilever, i.e., $\langle t=0 | a^+ a | t=0 \rangle = N_a$. With similar algebra to Eq. (8), we obtain

$$\frac{\rho(0)^2}{(1-4|\alpha_1(0)|^2)^{3/2}} - 1 = N_a . \tag{9}$$

Eqs. (8) and (9) result in $\rho(0) = \pm(N_a+1)^{-1/4}$ and $\alpha_1(0) = \pm\frac{1}{2}\sqrt{\frac{N_a}{(N_a+1)}}$, from which we can straightforwardly obtain the exact dynamics of the coupling system by solving Eqs. (4)-(7). Before doing the numerics, however, for a comparison, we would like to solve the Hamiltonian in Eq. (1) under the RWA by the same technique we used above. For the Hamiltonian

$$H = (\omega - i\Gamma_a)a^+a + (\nu - i\Gamma_b)b^+b - \kappa(ab^+ + a^+b), \tag{10}$$

where the counter-rotating-wave terms are removed due to the approximation that rapid oscillating terms resulted by large detuning between the two vibrational degrees of freedom can be effectively averaged out in the calculation in the case of weak coupling. If we still assume Eq. (2) to be the trial solution of the wavefunction, we have the Schrődinger equation

$$\begin{aligned}
i[\dot{\rho}(t) &+ \rho(t)\dot{\alpha}_1(t)a^+a^+ + \rho(t)\dot{\alpha}_2(t)a^+b^+ + \rho(t)\dot{\alpha}_3(t)b^+b^+]|A(t)\rangle \\
&= \rho(t)[2(\omega - i\Gamma_a)\alpha_1(t)a^+a^+ + (\omega - i\Gamma_a)\alpha_2(t)a^+b^+ \\
&\quad + (\nu - i\Gamma_b)\alpha_2(t)a^+b^+ + 2(\nu - i\Gamma_b)\alpha_3(t)b^+b^+ \\
&\quad - 2\kappa\alpha_1(t)a^+b^+ - \kappa\alpha_2(t)b^+b^+ \\
&\quad - 2\kappa\alpha_3(t)a^+b^+ - \kappa\alpha_2(t)a^+a^+]|A(t)\rangle,
\end{aligned} \tag{11}$$

which yields

$$i\dot{\rho}(t) = 0 \quad \rightarrow \quad \rho(t) = \rho(0) , \tag{12}$$

$$i\dot{\alpha}_1(t) = 2(\omega - i\Gamma_a)\alpha_1(t) - \kappa\alpha_2(t), \tag{13}$$

$$i\dot{\alpha}_2(t) = (\omega + \nu - i\Gamma_a - i\Gamma_b)\alpha_2(t) - 2\kappa\alpha_1(t) - 2\kappa\alpha_3(t), \tag{14}$$

$$i\dot{\alpha}_3(t) = 2(\nu - i\Gamma_b)\alpha_3(t) - \kappa\alpha_2(t). \tag{15}$$

Different from the case in the absence of the RWA, we can find analytical solutions from Eqs. (13) to (15). After the direct algebra shown in Appendix II, we have

$$\alpha_1(t) = \alpha_{10}^{(1)} e^{\Omega^{(1)}t} + \alpha_{10}^{(2)} e^{\Omega^{(2)}t} + \alpha_{10}^{(3)} e^{\Omega^{(3)}t},$$

$$\alpha_2(t) = \alpha_{20}^{(1)} e^{\Omega^{(1)}t} + \alpha_{20}^{(2)} e^{\Omega^{(2)}t} + \alpha_{20}^{(3)} e^{\Omega^{(3)}t}, \tag{16}$$

$$\alpha_3(t) = \alpha_{30}^{(1)} e^{\Omega^{(1)}t} + \alpha_{30}^{(2)} e^{\Omega^{(2)}t} + \alpha_{30}^{(3)} e^{\Omega^{(3)}t},$$

where

$$\alpha_{10}^{(1)} = -\frac{1}{2}\sqrt{\frac{N_a}{N_a+1}} \frac{(2\omega - 2i\Gamma_a - i\Omega^{(2)})(2\omega - 2i\Gamma_a - i\Omega^{(3)})}{(\Omega^{(3)} - \Omega^{(1)})(\Omega^{(2)} - \Omega^{(1)})},$$

$$\alpha_{10}^{(2)} = -\frac{1}{2}\sqrt{\frac{N_a}{N_a+1}} \frac{(2\nu - 2i\Gamma_b - i\Omega^{(2)})(2\omega - 2i\Gamma_a - i\Omega^{(3)})}{(\Omega^{(3)} - \Omega^{(2)})(\Omega^{(2)} - \Omega^{(1)})},$$

$$\alpha_{10}^{(3)} = -\frac{1}{2}\sqrt{\frac{N_a}{N_a+1}} \frac{(2\omega - 2i\Gamma_a - i\Omega^{(2)})(2\nu - 2i\Gamma_b - i\Omega^{(3)})}{(\Omega^{(3)} - \Omega^{(1)})(\Omega^{(3)} - \Omega^{(2)})},$$

$$\alpha_{20}^{(1)} = -\frac{1}{2}\sqrt{\frac{N_a}{N_a+1}} \frac{(2\nu - 2i\Gamma_b - i\Omega^{(1)})(2\omega - 2i\Gamma_a - i\Omega^{(2)})(2\omega - 2i\Gamma_a - i\Omega^{(3)})}{(\Omega^{(3)} - \Omega^{(1)})(\Omega^{(2)} - \Omega^{(1)})\kappa},$$

$$\alpha_{20}^{(2)} = -\frac{1}{2}\sqrt{\frac{N_a}{N_a+1}} \frac{(2\nu - 2i\Gamma_b - i\Omega^{(2)})(2\omega - 2i\Gamma_a - i\Omega^{(3)})(2\omega - 2i\Gamma_a - i\Omega^{(2)})}{(\Omega^{(3)} - \Omega^{(2)})(\Omega^{(2)} - \Omega^{(1)})\kappa}, \quad (17)$$

$$\alpha_{20}^{(3)} = \frac{1}{2}\sqrt{\frac{N_a}{N_a+1}} \frac{(2\nu - 2i\Gamma_b - i\Omega^{(3)})(2\omega - 2i\Gamma_a - i\Omega^{(3)})(2\omega - 2i\Gamma_a - i\Omega^{(2)})}{(\Omega^{(3)} - \Omega^{(2)})(\Omega^{(3)} - \Omega^{(1)})\kappa},$$

$$\alpha_{30}^{(1)} = \frac{1}{2}\sqrt{\frac{N_a}{N_a+1}} \frac{(2\omega - 2i\Gamma_a - i\Omega^{(2)})(2\omega - 2i\Gamma_a - i\Omega^{(3)})}{(\Omega^{(3)} - \Omega^{(1)})(\Omega^{(2)} - \Omega^{(1)})},$$

$$\alpha_{30}^{(2)} = -\frac{1}{2}\sqrt{\frac{N_a}{N_a+1}} \frac{(2\omega - 2i\Gamma_a - i\Omega^{(2)})(2\omega - 2i\Gamma_a - i\Omega^{(3)})}{(\Omega^{(3)} - \Omega^{(2)})(\Omega^{(2)} - \Omega^{(1)})},$$

$$\alpha_{30}^{(3)} = \frac{1}{2}\sqrt{\frac{N_a}{N_a+1}} \frac{(2\omega - 2i\Gamma_a - i\Omega^{(2)})(2\omega - 2i\Gamma_a - i\Omega^{(3)})}{(\Omega^{(3)} - \Omega^{(1)})(\Omega^{(3)} - \Omega^{(2)})}.$$

## III Numerical results

We consider an achieved cantilever with vibrational frequency 19.7 MHz [3] and assume the frequency of the ion's vibration to be 16 ~ 19.7 MHz. For simplicity, we fix the decay coefficients $\Gamma_a$, $\Gamma_b$ to be $\nu/1000$. But the coupling strength $\kappa$ will be changed in our calculation to check the difference between the treatments under the RWA and by the non-RWA. Other initial conditions are $\bar{n}_a(0) = N_a = 6$ and $\bar{n}_b(0) = 0$. In terms of our method presented in last section, no matter under which conditions, i.e., under the RWA or non-RWA, in order to obtain $\bar{n}_a(t)$ and $\bar{n}_b(t)$ by solving $|t\rangle$, we have to calculate $\alpha_1(t)$, $\alpha_2(t)$, $\alpha_3(t)$ and the initial conditions Eqs. (8) and (9).

When the two vibrational degrees of freedom are nearly resonant, i.e., $\omega \approx \nu$, the difference between the treatments under the RWA and with non-RWA is almost invisible in the case of very small coupling constants (i.e., $\kappa/\nu \leq 0.1$ in our calculation), as shown in Fig. 2. It can be well understood from the viewpoint of Jaynes and Cummings' seminal paper [17]: In this case, the counter-rotating terms are negligible and the RWA can be safely used. Our result also confirms the validity of the RWA employed in [10] and [14] because $\kappa/\nu \leq 0.1$ is satisfied there. Moreover, in this near resonant case, a weak coupling between the two vibrational degrees of freedom could achieve a quick cooling of the cantilever down to the ground state by a half coupling period, even in the case of relatively large decay coefficients $\Gamma_a$ and $\Gamma_b$.

To transfer the energy [10,14] or to make entanglement [10] with high fidelity, however, we prefer a large coupling $\kappa$. While with the increase of the coupling strength (e.g, $\kappa/\nu > 0.1$ in our calculation), the counter-rotating terms in the (near) resonant case is not negligible any more, and the RWA could not work well. This can be found in Fig. 3. To have a good transducer or to make high-quality quantum gates, we require the trapped ion to be decoupled from the cantilever at an exact time point. Therefore a non-RWA solution is essential to this situation.

With the increase of $\kappa$, more and more differences between the RWA and non-RWA solutions would appear. Particularly, if the two vibrational degrees of freedom are detuned, as shown in Figs. 4 and 5, the trapped ion cannot be a good prober and a cooler, because the phonon exchange is not complete between the ion and the cantilever. Both calculations with and without the RWA clearly demonstrate this point, while the non-RWA treatment could give a more accurate description of quantum behavior of the system.

## IV Discussion and conclusion

One point we should mention is that, we have only studied the first step in the proposal [14], because the second and the third steps have been well investigated by standard approaches. According to [14], after the vibrational quanta has been fully transferred to the ion, we switch off the coupling $\kappa$, and meanwhile turn on the red- or blue-detuned laser radiation. The current ion trap techniques have successfully coupled the internal and the vibrational degrees of freedom of the trapped ultracold ion [18]. But we have not yet found any experimental report for detecting the vibrational quanta of an ion beyond the sideband cooling regime. Considering an almost full phonon transfer from the cantilever to the ion, we choose $\bar{n}_a(0)$ to be 6 so that the vibrational quanta obtained by the ion can be detected at the level of a single quanta by means of the sophisticated optic method [15,16]. Nevertheless, from [14], it seems to be not difficult experimentally to detect a cantilever with the vibrational quanta $\bar{n}_a(0) \geq 4000$. As it in principle works for any $\bar{n}_a$ and $\bar{n}_b$, our method, providing an exact solution for the evolution of system, could be very helpful for the future experiment.

On the other hand, however, the phenomenological treatment of the decay in Eq. (1) due to thermal environment restricts our approach to work only on the cases of few vibrational quanta. Strictly speaking, if the vibrational quanta owned by the cantilever and by the trapped ion are not approaching zero, we cannot simply write the vibrational damping to be like in Eq. (1). Instead, we should utilize density matrix, not simply wave functions, to study the dynamics of the system [19]. Nevertheless, under the simple model with an ultracold ion coupled to a cantilever with only 6 vibrational quanta, Eq. (1) is valid in the case of $\max\{\Gamma_a, \Gamma_b\} \ll \kappa$. This approximation goes better when $\kappa/\max\{\Gamma_a,\Gamma_b\}$ is larger. In this sense, our investigation of the quantum dynamics of the system is exact only within the considered simple model with few vibrational quanta involved.

Besides, we have to mention again that Eq. (1) itself is an approximate description of the real coupling system without including three – and higher - order perturbative expansion terms of the Coulomb energy. So the model under consideration only works for the weak coupling cases. If we want to investigate a very strong coupling model, Eq. (1) should be modified to involve multi-phonon terms due to the higher-order terms of the perturbative expansion, which would be an interesting work in the future. What we have done here is to present an exact quantum behavior of the coupling system within the simple model in cases of relatively strong couplings (i.e., 1.9 MHz $\leq \kappa <$ 5.0 MHz in our solution).

To show the non-RWA effect experimentally, we need large coupling constant $\kappa$. As defined in [14], $\kappa \propto V_0/(mM\nu\omega d^6)^{1/2}$ where m and M are the mass of the ion and the cantilever, respectively, and $V_0$ and d are defined in Fig. 1. In terms of [3,14], the vibrational frequencies of the cantilever and the ion can be 19.7 MHz, and the available coupling constant $\kappa$ is about 0.3 MHz. So if we decrease the vibrational frequencies of both the cantilever and the ion by, for example, 4 times, $\kappa$ would be about 1.2 MHz, weaker than other two characteristic frequencies by only 4 times. As a result, the non-RWA effect studied in this paper would appear. Alternatively, the choice of a lighter cantilever, e.g., a single wall carbon nanotube, could also increase the coupling $\kappa$.

In summary, we have studied the time evolution of an electrostatically coupling system involving a trapped ultracold ion and a precooled cantilever, which is the first step of a previous proposal to employ the trapped ion as a transducer and to fast cool the cantilever down to its ground vibrational state. With the squeezed-state expansion technique, we can exactly demonstrate the quantum behavior of two coupled subsystems within the model given in [14], which is very different from the solution with the RWA under certain conditions. Since the field of QEMS is developing very quickly and is playing more and more important role in QIP for the scalability [10] and the qubit readout [11-13], we believe that our method as well as the results presented here could be useful in the fields of QEMS and QIP.

# IV Acknowledgements

This work is supported by National Natural Science Foundation of China under Grant Nos. 10474118 and 10274093, and by the National Fundamental Research Program of China under Grant Nos. 2001CB309309 and 2005CB724502.

## Appendix I

From the initial conditions $\alpha_2(0) = 0, \alpha_3(0) = 0$, and $\langle t=0 | t=0 \rangle = 1$, we have

$$|\rho(0)|^2 \langle 0 | e^{\alpha_1^*(0) aa} e^{\alpha_1(t) a^+ a^+} | 0 \rangle$$

$$= |\rho(0)|^2 \langle 0 | \sum_n \frac{1}{n!} [\alpha_1^*(0) a^2]^n \cdot \sum_m \frac{1}{m!} [\alpha_1(0)(a^+)^2]^m | 0 \rangle$$

$$= |\rho(0)|^2 \langle 0 | \sum_n \frac{1}{n!} [(\alpha_1^*(0))^n a^{2n}] \cdot \sum_m \frac{1}{m!} [(\alpha_1(0))^m (a^+)^{2m}] | 0 \rangle$$

$$= |\rho(0)|^2 \langle 0 | \sum_n \frac{1}{n!} [(\alpha_1^*(0))^n a^{2n}] \cdot \sum_n \frac{1}{n!} [(\alpha_1(0))^n (a^+)^{2n}] | 0 \rangle$$

$$= |\rho(0)|^2 \langle 0 | \sum_n \frac{1}{(n!)^2} [|\alpha_1(0)|^{2n} (2n)!] | 0 \rangle.$$

It is easy to prove

$$\sum_{n=0}^{\infty} \frac{1}{(n!)^2} [|\alpha_1(0)|^{2n} (2n)!] = \frac{1}{\sqrt{1 - 4|\alpha_1(0)|^2}}.$$

So we get $\dfrac{\rho(0)^2}{\sqrt{(1 - 4|\alpha_1(0)|^2)}} = 1$. Similarly, from $\langle t=0 | a^+ a | t=0 \rangle = N_a$, we can obtain

$$\frac{\rho(0)^2}{(1 - 4|\alpha_1(0)|^2)^{3/2}} - 1 = N_a,$$

where we have used

$$\sum_{n=0}^{\infty} \frac{1}{(n!)^2} [|\alpha_1(0)|^{2n} (2n+1)!] = \frac{\rho(0)^2}{(1 - 4|\alpha_1(0)|^2)^{3/2}}.$$

## Appendix II

For solving Eqs. (13) - (15), we set $\alpha_1(t) = \alpha_{10} e^{\Omega \cdot t}$, $\alpha_2(t) = \alpha_{20} e^{\Omega \cdot t}$ and $\alpha_3(t) = \alpha_{30} e^{\Omega \cdot t}$, and obtain

$$i\Omega \alpha_{10} = 2(\omega - i\Gamma_a) \alpha_{10} - \kappa \alpha_{20}$$

$$i\Omega \alpha_{20} = (\omega + \nu - i\Gamma_a - i\Gamma_b) \alpha_{20} - 2\kappa \alpha_{10} - 2\kappa \alpha_{30}$$

$$i\Omega \alpha_{30} = 2(\nu - i\Gamma_b) \alpha_{30} - \kappa \alpha_{20}.$$

By diagolizing the related determinant, we get to

$$\alpha_{20}^{(1,2,3)} = \frac{2(\omega - i\Gamma_a) - i\Omega^{(1,2,3)}}{\kappa} \alpha_{10}^{(1,2,3)},$$

$$\alpha_{30}^{(1,2,3)} = -\frac{2(\omega - i\Gamma_a) - i\Omega^{(1,2,3)}}{2(\omega - i\Gamma_b) - i\Omega^{(1,2,3)}} \alpha_{10}^{(1,2,3)},$$

where

$$\Omega^{(1)} = -[\Gamma_a + \Gamma_b + i(\omega + v),]$$

$$\Omega^{(2)} = -[\Gamma_a + \Gamma_b + i(\omega + v)] + \sqrt{-[i(\Gamma_a - \Gamma_b) - \omega + v]^2 - \kappa^2},$$

$$\Omega^{(3)} = -[\Gamma_a + \Gamma_b + i(\omega + v)] - \sqrt{-[i(\Gamma_a - \Gamma_b) - \omega + v]^2 - \kappa^2}.$$

By means of the initial conditions $\alpha_1(0) = \frac{1}{2}\sqrt{N_a/(N_a+1)}, \alpha_2(0) = 0, \alpha_3(0) = 0$, we obtain Eqs. (16) and (17).

### Captions of the Figures

**Fig.1** Schematic diagram of a single trapped ultracold ion coupled to a doubly clamped cantilever, where d is the separation comparable to the vibrational deviations of the ion and the cantilever from their equilibrium positions, and $V_0$ is the bias voltage.

**Fig.2** Time evolution of $\bar{n}_a$ and $\bar{n}_b$ calculated under and beyond the RWA, where ω=19.7 MHz, ν=19.7 MHz, κ=1.8 MHz, $\Gamma_a=\Gamma_b$=0.0197 MHz, and $\bar{n}_a(0) = 6, \bar{n}_b(0) = 0$.

**Fig.3** Time evolution of $\bar{n}_a$ and $\bar{n}_b$ calculated under and beyond the RWA, where ω=19.7 MHz, ν=19.7 MHz, κ=5.0 MHz, $\Gamma_a=\Gamma_b$=0.0197 MHz, and $\bar{n}_a(0) = 6, \bar{n}_b(0) = 0$.

**Fig.4** Time evolution of $\bar{n}_a$ and $\bar{n}_b$ calculated under and beyond the RWA, where ω=19.7 MHz, ν=16.0 MHz, κ= 4.0 MHz, $\Gamma_a=\Gamma_b$=0.0197MHz, and $\bar{n}_a(0) = 6, \bar{n}_b(0) = 0$.

**Fig.5** Time evolution of $\bar{n}_a$ and $\bar{n}_b$ calculated under and beyond the RWA, where ω=19.7 MHz, ν=16.0 MHz, κ=5.0 MHz, $\Gamma_a=\Gamma_b$=0.0197 MHz, and $\bar{n}_a(0) = 6, \bar{n}_b(0) = 0$.

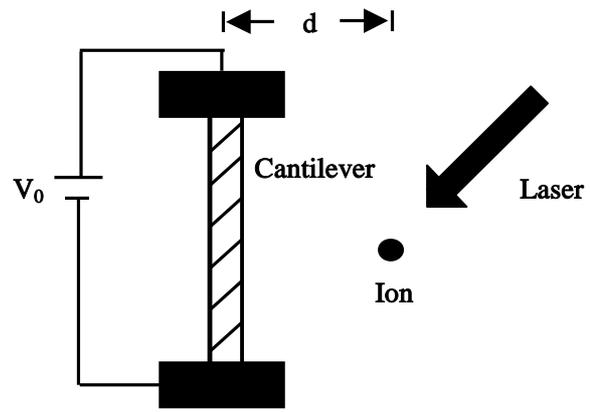

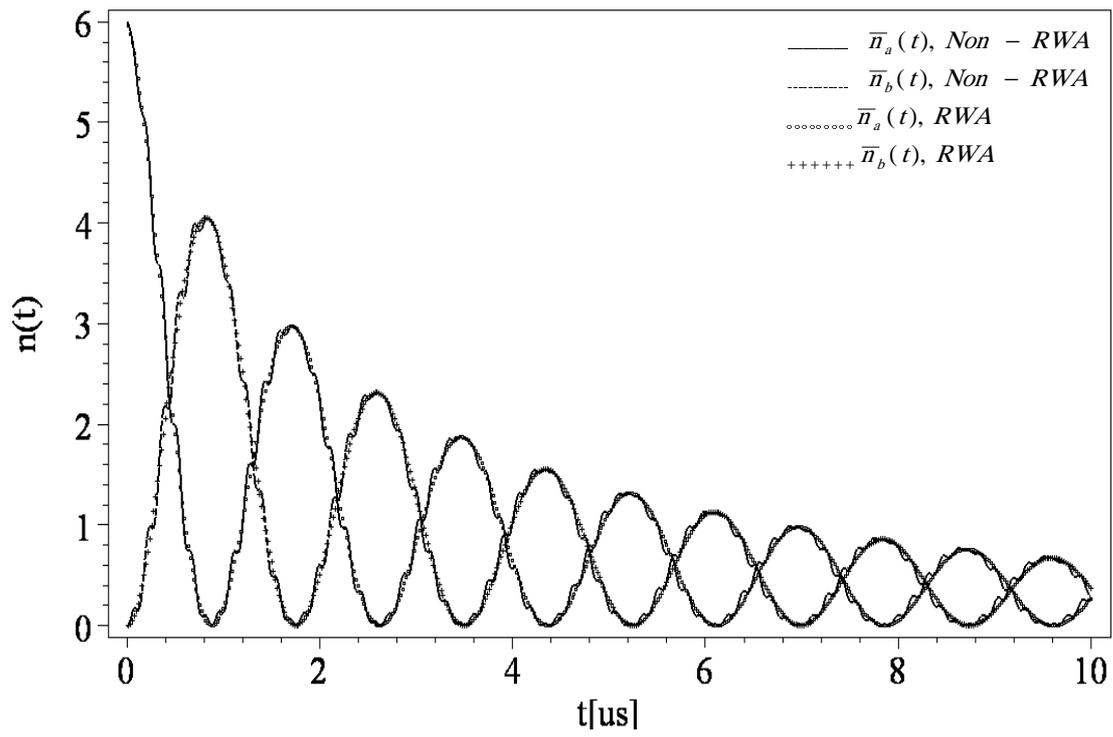

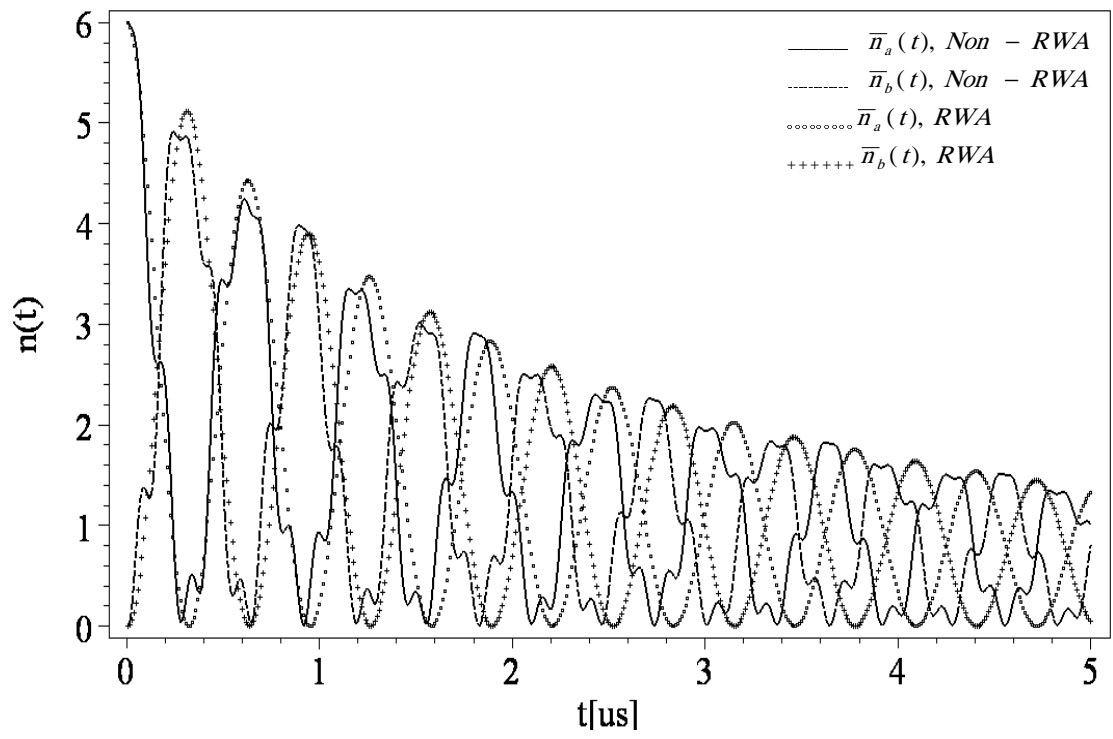

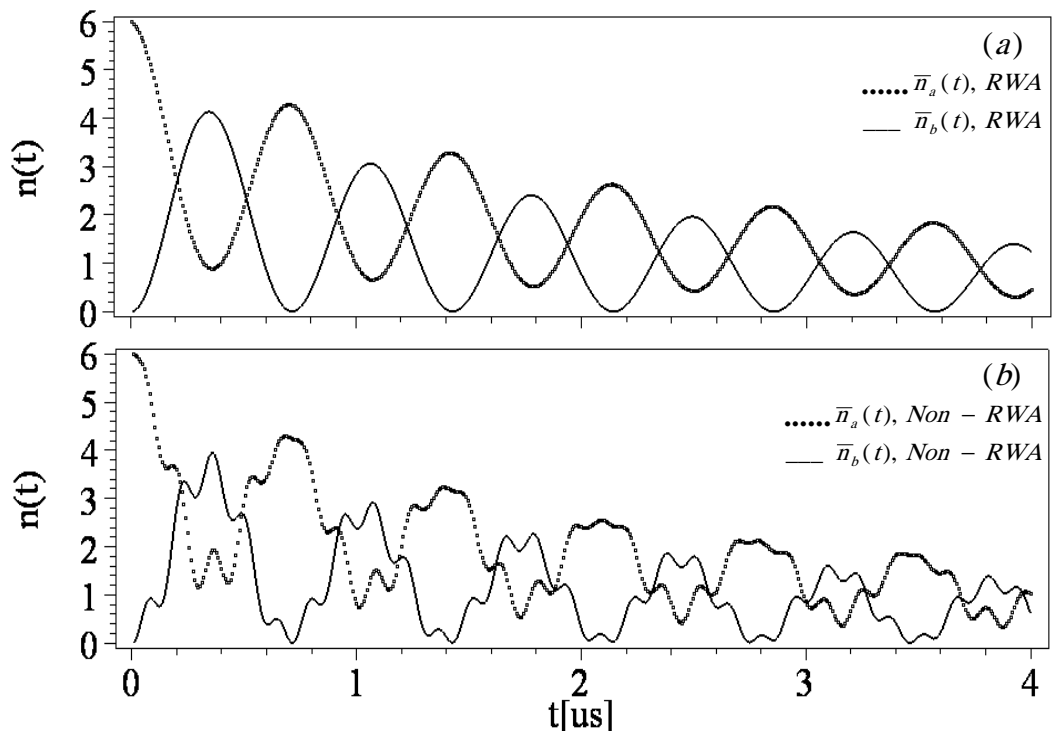

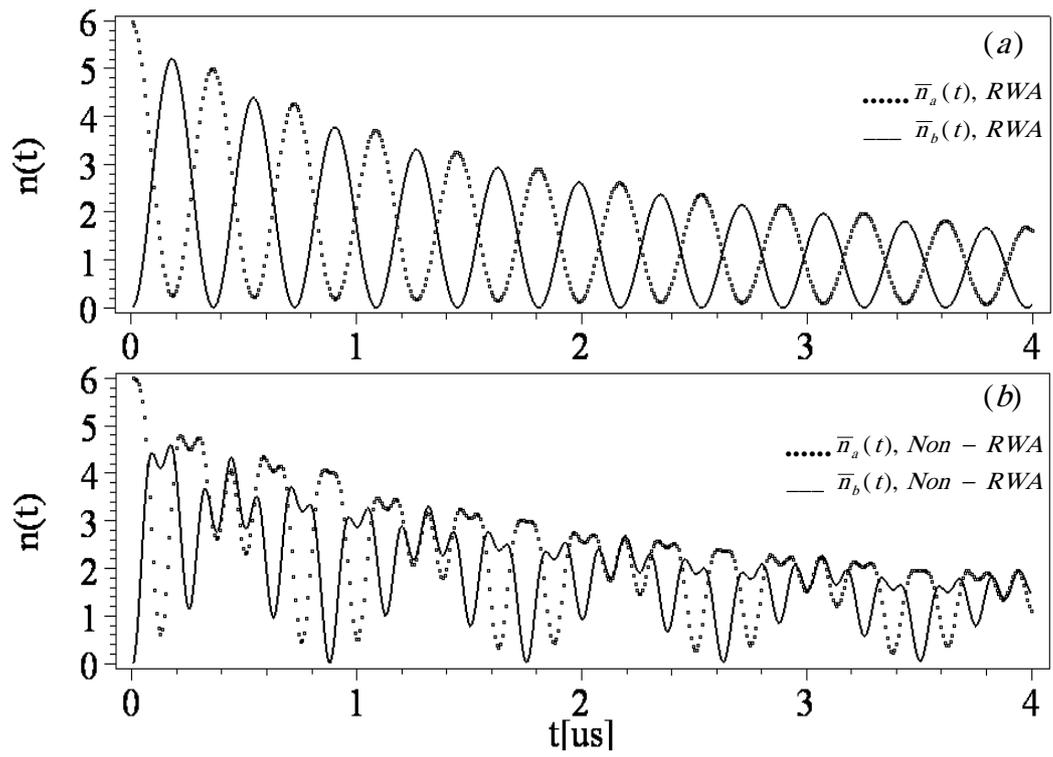